\documentclass{optica-article}

\journal{opticajournal} 

\articletype{Research Article}

\usepackage{lineno}
\nolinenumbers 
\usepackage{gensymb}
\usepackage{verbatim}

\begin{document}

\title{Alignment-Free Coupling to Arrays of Diamond Microdisk Cavities for Scalable Spin-Photon Interfaces}

\author{Helaman R. Flores,\authormark{1} Samuel R. Layton,\authormark{2}, Dirk Englund,\authormark{3}  and Ryan M. Camacho\authormark{1,*}}

\address{\authormark{1}Department of Electrical and Computer Engineering, Brigham Young University, 450 EB, Provo, UT 84604, USA\\
\authormark{2}Department of Mathematics, Brigham Young University, 275 TMCB, Provo, UT 84604, USA\\
\authormark{3} Department of Electrical Engineering and Computer Science, Massachusetts Institute of Technology, 77 Massachusetts Ave., Cambridge, MA 02139, USA}

\email{\authormark{*}camacho@byu.edu} 


\begin{abstract*} 
We propose a scalable design for a spin-photon interface to a color center in a diamond microdisk.  The design consists of a silicon oxynitride hexagonal lattice overlaid on a diamond microdisk to enable vertical emission from the microdisk into low-numerical aperture modes, with quantum efficiencies as high as 45\% for a tin vacancy (SnV) center. Our design is robust to manufacturing errors, potentially enabling large scale fabrication of quantum emitters coupled to optical collection modes. We also introduce a novel approach for optimizing the free space performance of a complex structure using a dipole model, achieving comparable results to full-wave finite difference time domain simulations with a 650,000 times reduction in computational time.

\end{abstract*}

\section{Introduction}

Efficiently interfacing optical fields with numerous quantum memories is a fundamental requirement for quantum networks. Solid-state systems, especially color centers in diamond, have emerged as promising candidates for such interfaces. They are well suited for building nodes in quantum repeater networks due to their ability to transmit quantum information via spin-entangled photons \cite{Ruf2021,Pompili2021,Orphal2023}. However, a significant limitation of these color centers, including Nitrogen-Vacancy (NV),  Silicon-Vacancy (SiV), and Tin Vacancy (SnV) centers, is their suboptimal emission efficiency into preferred spatial and spectral modes, particularly into the zero-phonon line (ZPL) \cite{duan21,Humphreys2018-pj}. Enhancing the photonic emission of color centers in the ZPL and effectively coupling these photons into network channels (such as fibers or free space) is crucial for improving entanglement generation rates, a current bottleneck in quantum network performance \cite{Humphreys2018-pj}.

For spin-photon interfaces in such systems, the figure of merit, denoted as \(\eta\), represents the probability that an excited state of the emitter leads to a detected photon in the ZPL. This efficiency, \(\eta\), can be conceptualized as a product of two efficiencies: \(\eta_{ZPL}\), the proportion of emissions into the ZPL, and \(\eta_{col}\), the collection efficiency of these emissions into the desired mode \cite{duan21}. Optimizing both these components is vital for maximizing \(\eta\), thereby enhancing the interface efficiency essential for scalable and efficient quantum networks.


We define the figure of merit \(\eta\) as the probability that an excited state of the emitter results in a photon detected in the zero-phonon-line (ZPL),

\begin{align}
\eta = \eta_{ZPL}\eta_{col},
\end{align}

where \(\eta_{ZPL}\) is the fraction of emission of the color center into the ZPL \cite{li15}, and \(\eta_{col}\) represents the fraction of emission coupled into the desired output mode \cite{gould16}. The factor \(\eta_{ZPL}\) can be determined using the Purcell enhancement \(F\) of the device at the targeted wavelength, and depends on the emission ratios \(\Gamma _{\textrm {total},0}\) and \(\Gamma _{\textrm {ZPL},0}\), which represent the total unmodified emission rate and ZPL emission rate of the chosen color center, respectively \cite{duan21}:

\begin{align}
\eta_{ZPL} = \frac{F}{F+(\frac{\Gamma_{\textrm{total},0}}{\Gamma_{\textrm{ZPL},0}}-1)}.
\end{align}

The parameter \(\eta_{col}\) is defined as the radial power in the far field collected into numerical apertures (NA) less than 0.7. Prior research has demonstrated that both \(\eta_{ZPL}\) and \(\eta_{col}\) can be simultaneously enhanced by positioning a color center in specially designed optical cavities \cite{gould16,duan21}. For silicon vacancy (SiV) centers, simulations predict free-space collection quantum efficiencies up to 41\% \cite{duan21}. However, fabricating such devices necessitates intricate diamond etching, a process challenged by diamond's material properties. Empirical evaluations of individually designed devices have shown average quantum efficiencies up to 5.5\% for nitrogen vacancy centers \cite{gould16}. Recent studies suggest that SnV centers may be more advantageous for quantum networking applications \cite{Bradac2019-hl, trusheim2020-mt, Takou:21, PhysRevLett.129.173603-Trusheim, PhysRevX.13.031022-Rosenthal}, with \(\Gamma _{\textrm {total},0}/\Gamma _{\textrm {ZPL},0}-1 \approx 0.25\), and are thus employed for computations in this work.

Here, we introduce a design for the vertical emission of spin-polarization entangled photons from diamond color centers, achieving quantum efficiencies up to 45\% for a SnV center. This approach mitigates the precision requirements for diamond etching, leading to a design that is more scalable and tolerant of manufacturing variances.

Additionally, we present a numerically efficient optimization strategy for this multilayered diamond emission device. Our approach yields results comparable to those from a comprehensive full-wave finite difference time domain (FDTD) parameter sweep, which can be computationally intensive, particularly for resonant structures where electric fields remain trapped in the cavity for extended durations \cite{OSKOOI2010687}. We propose a novel method for designing emitter devices that uses a simplified dipole representation of scattering from a near-field perturbation \cite{duan21,Zhu:13}, coupled with a single control simulation performed using FDTD. This method facilitates rapid optimization of key parameters in complex designs, and may be applicable to photonic device design more generally.

\section{Design}
\begin{figure}[ht]
\centering\includegraphics[width=10cm]{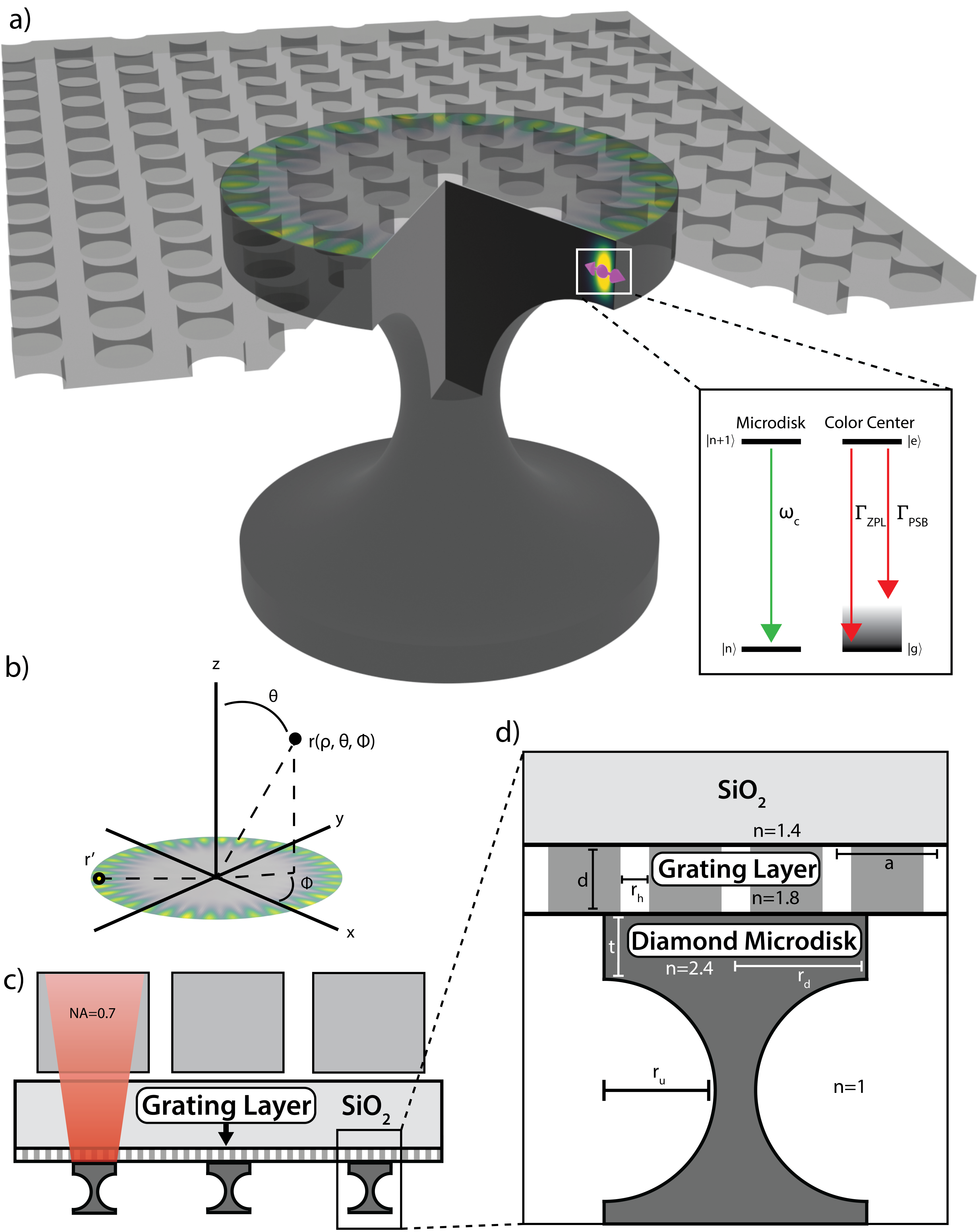}
\caption{Design concept. (a) Perspective from above disk, with a 60\textdegree  cutout to show the cross-section and bulk oxide removed. The purple mark represents the point dipole source used to excite the disk mode. (b) Coordinate system for the far field emission analysis (Eqs 9-10). A dipole location is highlighted in yellow, and a point in the far field in black. (c) $x-z$ cross-section of a microdisk array couple to numerical apertures. Disks would most likely be fabricated with an undercut \cite{mitchell19}, part of which is included in the FDTD simulations. (d) Microdisk with design parameters labeled.}
\label{fig:design}
\end{figure}

 Microdisk resonators have high quality factors, and thus provide the Purcell enhancement necessary for $\eta_{ZPL}$ to approach unity. However, radiation loss is dominated by radial emission in the plane of the disk. In order to achieve out-of-plane emission in the vertical direction, and thus high \(\eta_{col}\), our design builds on previous work showing that carefully placed perturbations in the near field mode of the microdisk can scatter photons vertically instead of radially. The difficulty of accurately etching the necessary patterns into diamond (n=2.4) prompts the exploration of alternative device designs and materials \cite{mitchell19}. To avoid etching the diamond more than necessary to create resonance, we propose a design in which a grating layer of silicon oxynitride (Si\(_{3}\)O\(_x\)N\(_{y}\)) is etched with a triangular lattice of holes and then placed on top of an array of fabricated diamond microdisks, as shown in \ref{fig:design}. A bulk silicon dioxide glass substrate (n=1.4) supports the grating layer, also shown in Figure \ref{fig:design}. The lattice holes scatter light from the microdisk mode so as to constructively interfere within the target numerical aperture.

\begin{figure}[ht]
\centering\includegraphics[width=6cm]{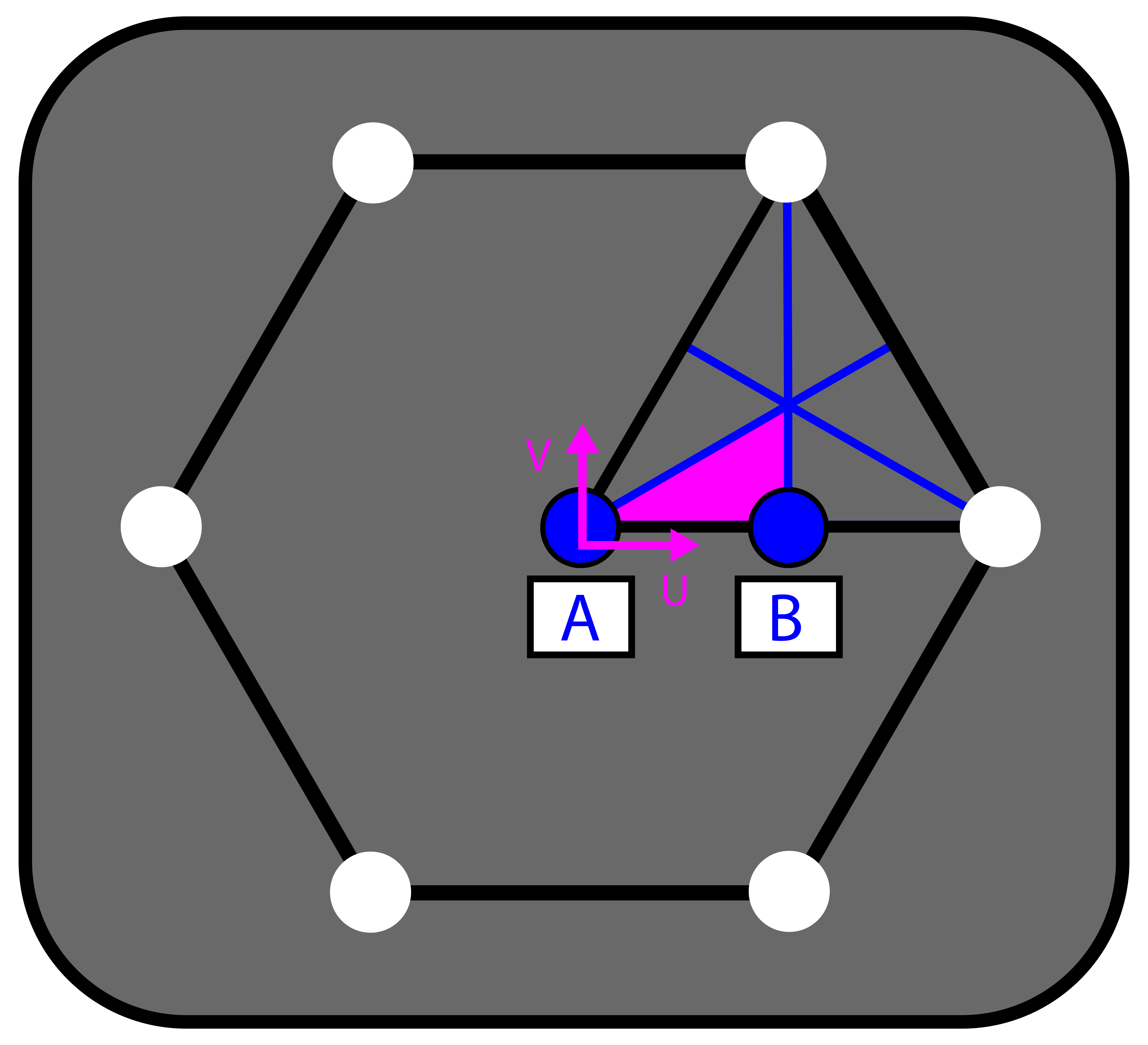}
\caption{Hexagonal lattice unit cell. The blue symmetry lines identify a reduced symmetry area for \(u\) and \(v\), highlighted in pink. Points A and B are key alignment points (\(u, v\)) exhibiting high radial symmetry.}
\label{fig:hex_cell}
\end{figure}

One pivotal aspect of our design is the capacity to overlay the grating layer onto the microdisk while accommodating reasonable alignment tolerances between the grating holes and the microdisk. This alignment is defined by a vector \((u, v)\) in the \(x-y\) plane, signifying the relative position of a lattice hole's center to the disk's center. Due to the hexagonal symmetry of the lattice, we can constrain the domains of \(u\) and \(v\) to \(0 \leq u \leq a/4\) and \(0 \leq v \leq u\tan(\pi/6)\), as illustrated in Fig. \ref{fig:hex_cell}.

Additionally, fabrication variances will inevitably lead to disparities in the actualized devices. Consequently, we aim for a design that exhibits robustness to these variations, focusing particularly on parameters such as \(a\), \(r_h\), \(t\), and \((u, v)\). In the design proposed, the utilization of a large microdisk array against a singular triangular lattice facilitates vertical free-space emission, even without precise alignment of the layers.

\begin{figure}[ht]
\centering\includegraphics[width=12cm]{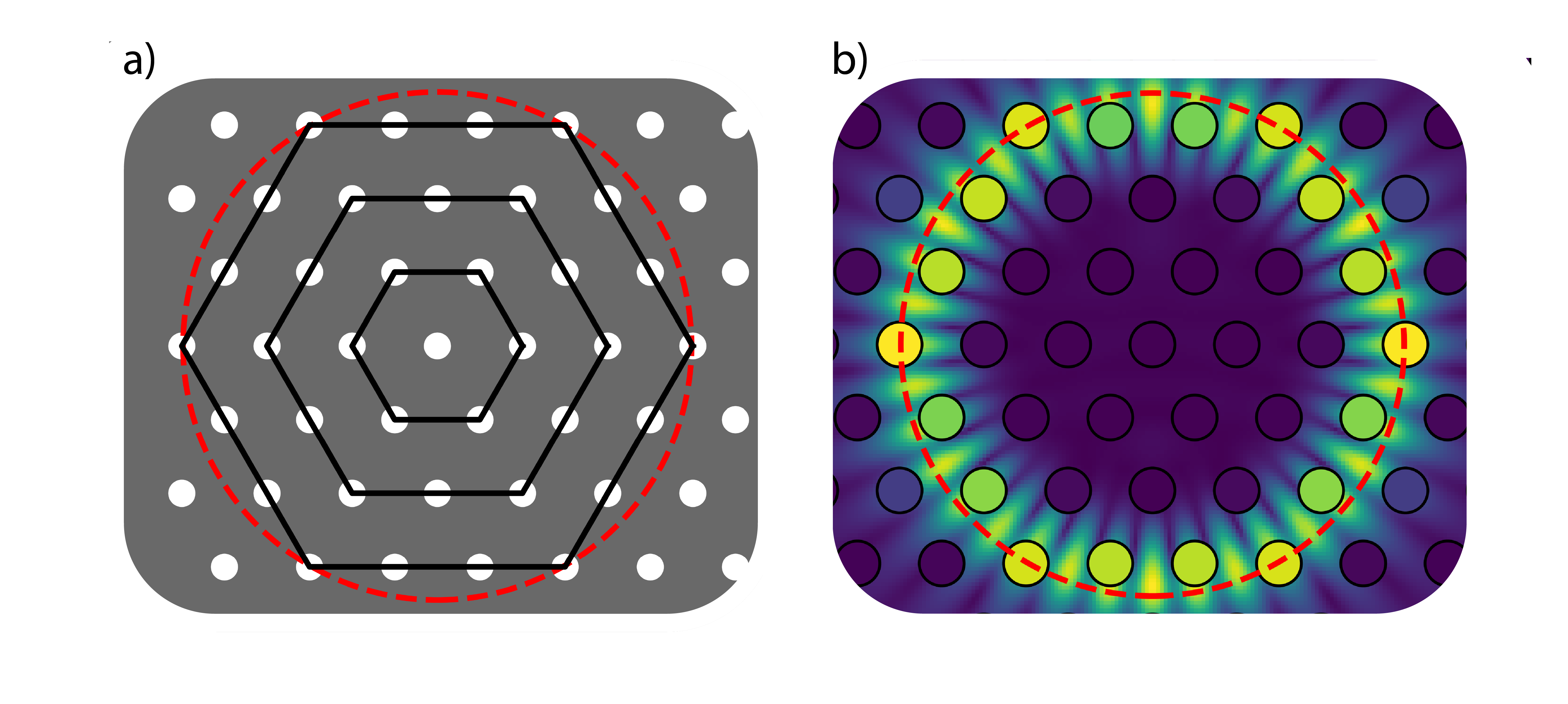}
\caption{ (a) Expanding hexagonal traces from a center lattice point in a triangular lattice.  For $n = 3$, 18 lattice holes form the perimeter of the hexagon.  Two-thirds (12) of these holes are equidistant from the center, and the circumference of the microdisk is circled in red. (b) Overlap between the lattice holes and the optical mode of the bare microdisk, with the circumference circled in red. The mode is calculated in FDTD without the lattice etched in the grating layer. Holes are highlighted in proportion to the electric field magnitude incident at their center. }
\label{fig:m18choice}
\end{figure}

To narrow the parameter space for the lattice constant \(a\), we compare the overlap between a given lattice and the mode of the microdisk. In a triangular lattice, the perimeter of the $n$th regular hexagon centered on any lattice point will intersect with $6n$ other lattice points, as shown in Fig. \ref{fig:m18choice}.   If the center of the microdisk is aligned with a lattice point and radius of the disk $r_d$ and the lattice constant $a$ are chosen properly, the disk mode will strongly overlap with one of these hexagons.

Based on previous work demonstrating that perturbations in a microdisk structure can be modeled as an array of dipole scatterers \cite{duan21, Zhu:13}, constructive interference into lower numerical apertures will occur when the number of equally spaced azimuthal scatterers is close to the azimuthal mode number $m$ of the disk \cite{duan21, xie2023}. We choose a disk mode of \(m=18\) to couple to hexagonal trace 3 in a triangular lattice, shown in Fig. \ref{fig:m18choice}. This choice has the distinct advantage of a having a high ratio of equidistant holes from the center (2/3) in the target hexagonal trace, creating a consistent overlap with the concentrated electric field at the edge of the disk when \(u,v << a\). For the \(m=18\) mode, \(r_d\) is also large enough to maintain a high Purcell enhancement.

\section{Simulation and Modeling}

To decrease simulation time, we apply a dipole model as an efficient alternative to computationally expensive finite-difference time-domain (FDTD) parameter sweeps. This approach is grounded in existing research indicating that refractive index variations in the vicinity of a microdisk's near field can effectively be represented as dipole scatterers \cite{duan21,Zhu:13}. 

The emission characteristics of the structure are derived through far-field projections from electromagnetic currents captured on a two-dimensional surface \(s\). These projections are facilitated by calculating the magnetic (\(A\)) and electric (\(F\)) vector potentials, derived from the surface electric (\(J_s\)) and magnetic (\(M_s\)) currents \cite{john10}:

\begin{equation} A(R)=\mu \oint_s J_s(r_c)\frac{e^{-jk|r-r_c|}}{4\pi|r-r_c|}ds' \end{equation}

\begin{equation} F(R)=\epsilon \oint_s M_s(r_c)\frac{e^{-jk|r-r_c|}}{4\pi|r-r_c|}ds'. \end{equation}
where \(r\) is the location of the point in the far field and \(r_c\) is the location of the currents. 

The electric far field \(E_{ff}\), magnetic far field field \(H_{ff}\) and far field poynting vector \(S_{ff}\) may then be calculated as \cite{john10}

\begin{equation} E_{ff} = -j\omega[A+\frac{1}{k^2}\nabla(\nabla \cdot A)] - \frac{1}{\epsilon}\nabla \times F, \end{equation}

\begin{equation} H_{ff} = -j\omega[F+\frac{1}{k^2}\nabla(\nabla \cdot F)] - \frac{1}{\mu}\nabla \times A, \end{equation}

\begin{equation}
    S_{ff} = E_{FF} \times H_{ff},
\end{equation}

where \(\mu\) and \(\epsilon\) represent the permeability and the permittivity of the medium, respectively. The angular frequency \(\omega\) and wave vector \(k\) are calculated based on the ZPL emission wavelength.   

We now examine the dipole approximation to this equation. The fields collected by our FDTD simulations will appear to have been sourced by dipoles centered on refractive-index perturbations. This is because holes in the grating layer cause large scattering in comparison to the natural losses of a microdisk with high a quality factor. Approximating each hole as a Hertzian dipole source of length $l$ leads to a magnetic vector potential of an array of Hertzian dipoles \cite{Ulaby_Ravaioli_2023}:

\begin{equation} A(R)=\frac{\mu}{4\pi}l\sum\frac{e^{-jk|r-r'_n|}}{|r-r'_n|}[I_{nx} \hat{x} + I_{ny} \hat{y} + I_{nz} \hat{z}],\end{equation}

where \(r_n\) is the position of a point in the far field and \(r'_n\) is the position of the dipole. In this equation, each dipole is represented as a sum of three excitation currents along Cartesian directions. The dipole excitation current \(I_{nx,ny,nz}\) is a coefficient proportional to the complex electric field incident at the location of the hole in the lattice, which can be obtained from an FDTD simulation. However, obtaining a specific near-field for every possible lattice constant and alignment is computationally expensive.  We therefore propose a simplified model in which the electric field for all possible lattice parameters is approximated by a single near-field calculation of the microdisk in which no lattice holes are present. To calculate the far-field profile of a particular lattice geometry, we find the phase and amplitude for \(I_{nx,ny,nz}\) in the simplified near field at the center of each lattice hole. The corresponding far field for the layered microdisk structure can then be modeled by

\begin{equation} 
\begin{aligned}
E_{ff} = \frac{\eta lk}{4\pi j}\sum_{n=0}^{N}\frac{e^{-jk|r-r'_n|}}{|r-r'_n|}[\hat{\theta}(I_{nx}cos(\theta)cos(\phi)\\+I_{ny}cos(\theta)sin(\phi)-I_{nz}sin(\theta))\\+\hat{\phi}(-I_{nx}sin(\phi)+I_{ny}cos(\phi))]
\end{aligned}
\end{equation}

\begin{equation}
    S_{ff} = \frac{|E_{ff}|^2}{\eta}\hat{r},
\end{equation}
where \(\theta\) and \(\phi\) are defined as the angles between the far field point and the origin, and \(\eta\) is the impedance of light in glass. We assume the dipole length \(l\) is the same for every lattice hole, and the wave vector \(k\) is the spatial frequency of light in the glass substrate.

In FDTD and with the simplified dipole model,  we sum up the poynting vector over a sphere centered on the origin with a large enough radius such that \(r<<r'\). For the dipole model, we sum the power collected in the glass by a numerical aperture of 0.7 and normalize it by the total power passing through the sphere in order to obtain \(\eta_{col}\). In FDTD, the surface currents used in the far-field projection must be evaluated in a homogeneous medium. To apply this appropriately to our layered microdisk structure, we project the surface currents collected by monitors placed only in the glass substrate. This will capture light escaping from the edge of the microdisk to around \(\theta=70\degree\). We then normalize the resulting far-field projection by the ratio of the power through the far field monitors to the total power that exits the simulation.  The FDTD near and far fields are calculated using commerical software\cite{Lumerical}, and Eqs. 9-10 of the dipole model are implemented in Python. 

We note that the simplified dipole model matches more accurately with emission profile of FDTD simulations when we assume that \(I_{nz}=0\). This may be attributed to the fact that, owing to boundary conditions, light traveling in the plane of the lattice and polarized in the out-of-plane \(\hat{z}\) direction will not experience a discontinuity anywhere along the boundary of a lattice hole, leading to reduced scattering relative to the in-plane components of the near field.  

\begin{figure}[ht]
\centering\includegraphics[width=12cm]{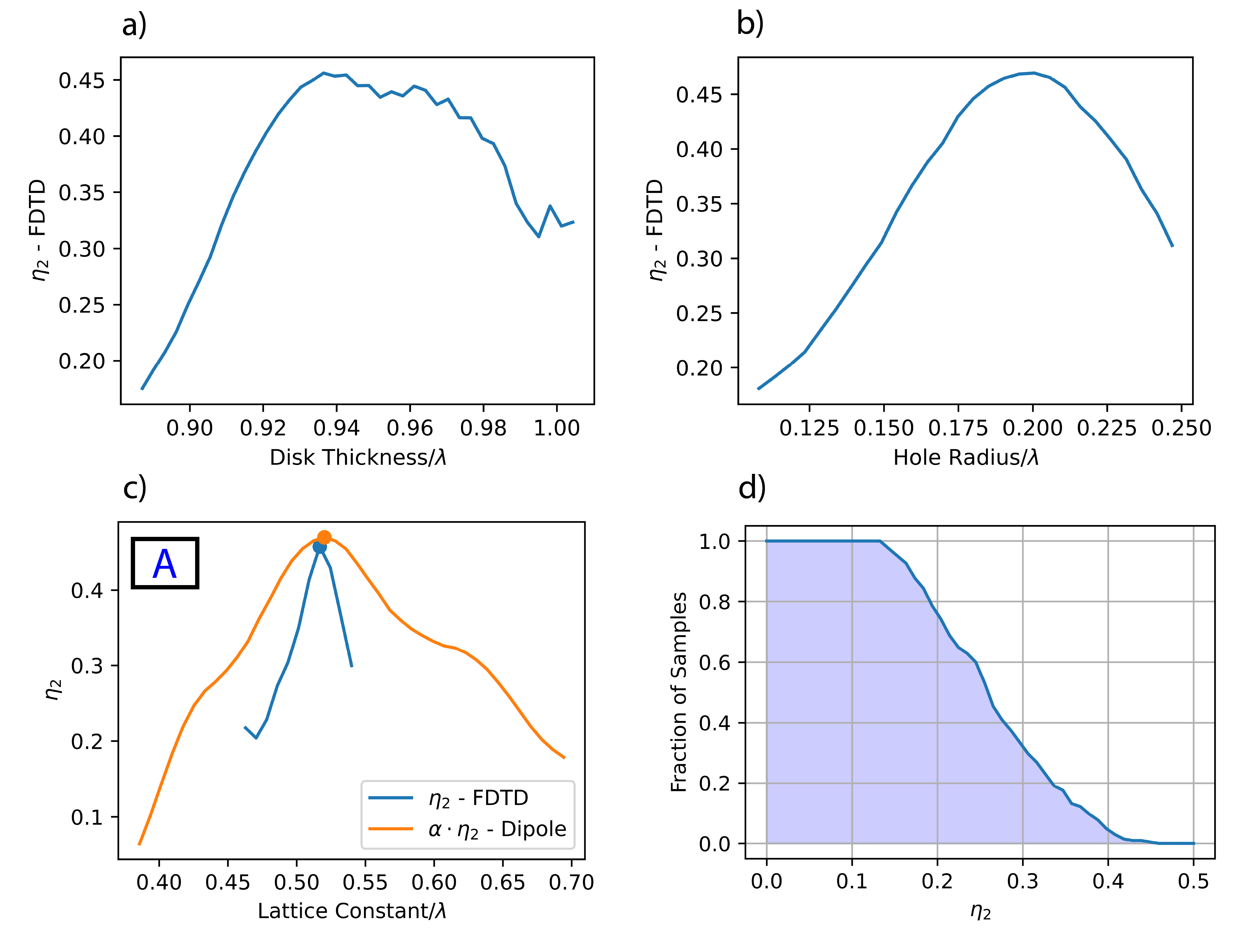}
\caption{ Manufacturing robustness according to the following parameters: (a) Disk thickness \(t\) from FDTD simulation. (b) Hole radius \(r_h\) from FDTD. (c) Lattice constant \(a\), with results from the dipole model plotted alongside FDTD. Peak performance is denoted by a dot. (d) Cumulative fraction of samples above \(\eta_{col}\) for random $(u, v)$ alignments sampled from FDTD. }
\label{fig:robustness}
\end{figure}

A single emission pattern can be calculated using the simplified dipole model in 3.6 s on one core of a Intel® Core™ i7-10750H processor.  In contrast, evaluating a single emission pattern in FDTD requires 5.1 hours on a cluster of 128 cores of AMD EPYC 7763 processors.  The simplified model may therefore be used to search a much larger parameter space to identify optimal emission patterns.  FDTD simulations may then be run to verify the pattern and optimize if necessary.
The simplified dipole model predicts an optical lattice constant \(a=0.524/\lambda\), which is consistent with FDTD simulations, as shown in Fig. \ref{fig:robustness}.  

Table \ref{tab:optparams} shows the parameters used to obtain the optimal far-field emission.  With a Purcell enhancement of 52.6 and \(\eta_{col}=0.47\), we find an upper \(\eta\) of 46\% for a SnV center. We also list in Table 2 the fraction of light naturally emitted into the ZPL for various color centers and the associated spectral efficiency \(\eta_{ZPL}\).

\begin{table}
    \centering
    \begin{tabular}{|c|c|c|c|c|c|c|c|c|} 
    \hline
        Parameters &\(a\) & \(d\) & \(r_h\) & \(r_d\) & \(t\) & \(r_u\) & \(u\) & \(v\) \\
     \hline
        Units of \(\lambda_0\) &\(0.5168\) & \(0.2931\) & \(0.2\) & \(1.5427\) & \(0.9411\)  &\(1.45\) & \(0\) & \(0\)\\
    \hline
    \end{tabular}
    \caption{Parameters of the optimized layered microdisk obtained from FDTD}
    \label{tab:optparams}
\end{table}

\begin{table}
    \centering
    \begin{tabular}{|c|c|c|c|} 
    \hline
        Color Center & NV & SiV & SnV \\
     \hline
        Estimated ZPL Emission & 0.03 & 0.7 & 0.8 \\
    \hline
        Spectral Efficiency \(\eta_{ZPL}\) & 0.62 & 0.99 & 0.995\\
    \hline
    \end{tabular}
    \caption{Spectral Efficiency for various color centers in diamond. }
    \label{tab:zpletas}
\end{table}

\begin{figure}[ht]
\centering\includegraphics[width=11cm]{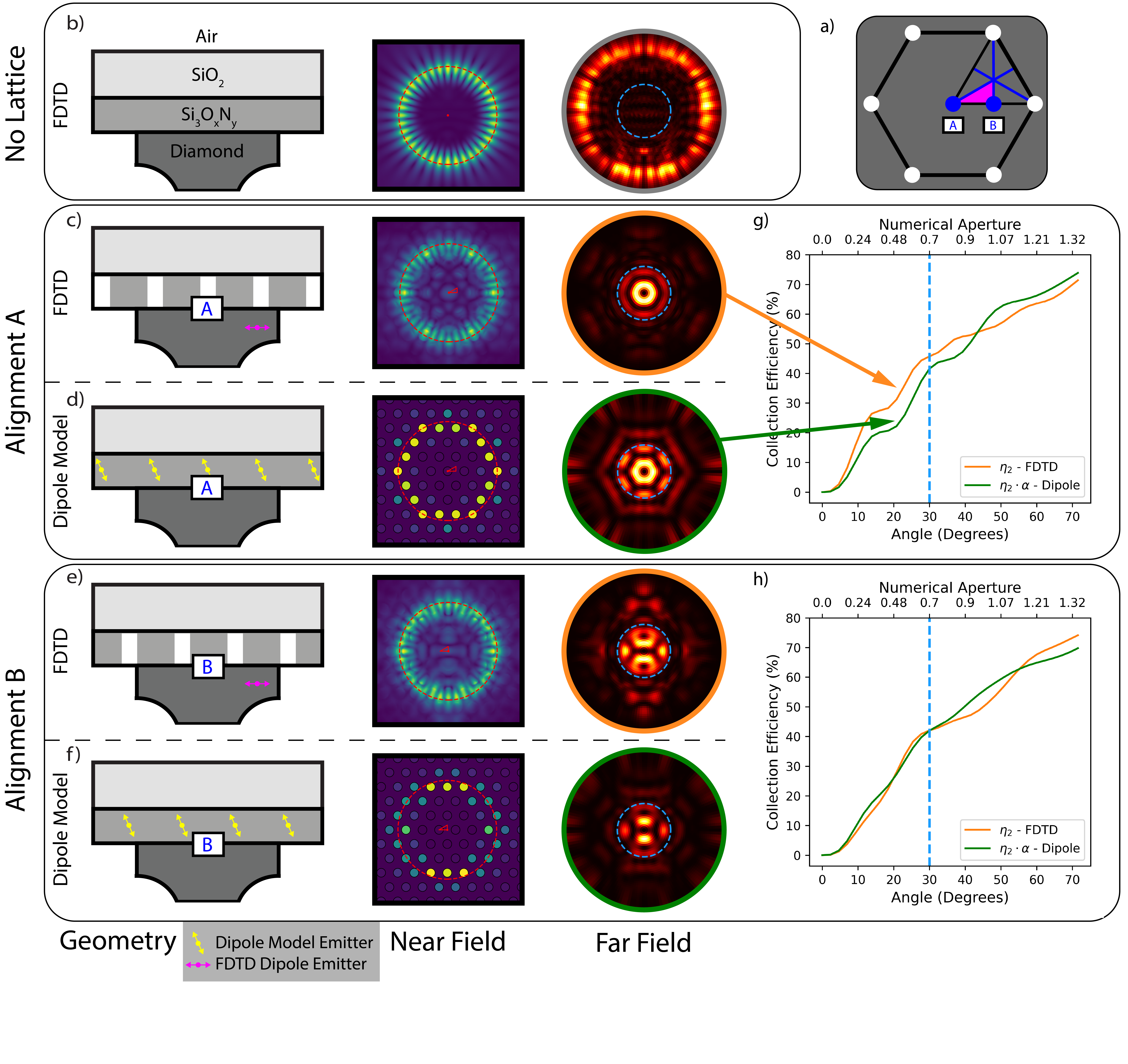}
\caption{ Comparison of emission of the optimized layered microdisk using FDTD and the simplified dipole model. Near fields are displayed inside the grating layer, and the far field in the upper hemisphere is depicted for each model. a) High-symmetry points A and B in a hexagonal lattice. b) FDTD simulation omitting the etch in the grating layer. The dashed red line depicts the boundary of the microdisk. The near field inside of the grating layer is used to find \(I_{nx,ny,nz}\) in the dipole model. Without scatterers, the farfield is directed radially away from the microdisk. c) FDTD simulation with the disk center aligned with point A on the lattice. FDTD near field and resulting far field. d) Geometry of the dipole model with the center of the disk aligned on point A. Dipole near field and resulting far field. e) FDTD simulation with the disk center aligned over point B on the lattice. FDTD near field and resulting far field. f) Geometry of the dipole model with the center of the disk aligned on point B. Dipole near field and resulting far field. g) Comparison of collection efficiency (\(\eta_{col}\)) as the target numerical aperture increases for alignment A according to FDTD simulation and the dipole model h) Comparison of collection for alignment B. }
\label{dipolevsfdtd}
\end{figure}

The best correspondence between the dipole scattering model and FDTD simulation occurrs when the center of the disk is aligned over symmetry points A and B on the lattice, as shown in Fig. \ref{dipolevsfdtd}. The correspondence between the calculated far-field emission patterns of the two models is reduced when the alignment is not near one of these symmetry points.  This may be because each simulation in the dipole model uses the near field mode of a bare disk with no grating to calculate \(I_{nx,ny,nz}\), which is then used in Eq. 9 to calculate the far field.  When the lattice and the microdisk are not aligned at a high-symmetry point, the non-uniformites in the near-field caused by the lattice holes make this approximation less valid.  Nonetheless, we find by numerical experimentation that optimization near the high symmetry points using the simplified dipole model is sufficient to optimize the far-field verified by FDTD even away from these high-symmetry points.  

Another physical effect missing from the dipole model is the consequence of the refractive index differences between the layers.  The difference between the diamond microdisk and the silicon dioxide lattice is much less than the difference between the disk and the air below it, leading to preferential emission into the bulk silicon dioxide. To include this effect in our model, we phenomenologically multiply all calculated far-field emission patterns for a given geometry by a constant parameter \(\alpha\). Because the monitors in FDTD simulation are placed to capture up to a 70\textdegree  cone of light exiting the simulation region from the microdisk, we perform a least-squares fit between the dipole model and the FDTD model up to 70\textdegree to find \(\alpha\), which has a  root mean squared error of \(0.05\) and \(0.03\) for symmetry points A and B respectively. 

It is notable that the dipole model demonstrates reasonable accuracy even  without accounting for variations in parameters such as disk thickness \(t\), hole radius \(r_h\), lattice depth \(d\), and disk radius \(r_d\). As noted above, the model's primary utility is in rapidly searching for optimal values of the lattice constant $a$. Initialization of the dipole model requires just one FDTD simulation to calculate the microdisk's mode, influenced by \(d\), \(r_d\), and \(t\). This helps isolate the lattice constant as the key variable for optimization.  It accurately identifies the optimal lattice constant with a precision of approximately 1.5\%, demonstrating the model's potential in streamlining the design process for complex photonic structures like the layered diamond microdisk resonator.

To demonstrate the robustness of the optimized design to fabrication variations, we swept parameters such as lattice constant, hole radius, and disk thickness and monitored the decrease in efficiency. Statistical sampling of 205 potential random variations indicates that the design consistently yields high performance, as shown in Figure \ref{fig:robustness}. A significant percentage of samples (~5\%) maintain far field efficiencies above 40\%, and the average Purcell enhancement is 49.5.  Experimentally, many color centers can be simultaenously implanted in a diamond sample and only those with desirable spectral properties may be addressed, which may increase the yield to near unity for this design.

Figure \ref{fig:robustness} illustrates the robustness to manufacturing variations, showing the impact of different parameters. Panel (a) presents results from FDTD simulations for disk thickness \(t\), while panel (b) shows the effects of variations in hole radius \(r_h\). Panel (c) compares the optimizations of the lattice constant \(a\) from both the dipole model and FDTD simulations.  Panel (d) shows the cumulative fraction of samples exceeding a certain \(\eta_{col}\) threshold for 205 random \((u, v)\) alignments, as sampled from FDTD simulations.

\section{Discussion}

The layered microdisk resonator presented here achieves a high coupling efficiency (\(\eta_{col}\)) of 0.46, attributed to the oxynitride grating layer that enhances vertical emission. This layer directs most far-field emission into the upper hemisphere. However, the collection efficiency (\(\eta_{col}\)) is capped due to constructive interference into higher numerical apertures, a consequence of the symmetric lattice structure. Optimizing hole locations in the nitride layer could enhance lower numerical aperture interference, but would require precise fabrication alignment.

Our focus on light collection into a specific numerical aperture, like a cryogenic objective, might differ from scalable designs involving microdisk arrays coupling into optical fibers. In such designs, a spatial light modulator could adjust free-space collection to fiber mode coupling, with grating hole arrangements tailored for Gaussian beam profiles.

Our control FDTD simulation-based dipole model accurately predicts the optimal lattice constant to within 5 nm. This model's potential extends to designing devices needing free-space coupling from near-field modes, though it currently does not fully account for asymmetric perturbations, preferential upper hemisphere emission, and lattice reflections. Refinements, such as modeling scattering at hole edges, could enhance accuracy, making this approach a valuable tool for solving various resonator scattering problems and reducing dependence on extensive FDTD simulations.

\section{Conclusion}


We presented a scalable alignment-free design for free-space coupling from diamond microdisks, offering a solution that alleviates the challenges of precision nanofabrication in diamond. This design demonstrates robustness across several parameters: the thickness of the diamond microdisk, the alignment of the grating and microdisk layers, the lattice constant of the nitride layer, and the dimensions of etches in the nitride.  We also developed a rapid optimization method using a single FDTD simulation and a dipole model instead of comprehensive full-wave parameter sweeps. This method significantly accelerates the design process for complex structures, especially for resonant devices with high quality factors. The resulting optimized layered microdisk design allows for creating arrays of quantum emitters with high alignment yields and a maximum quantum efficiency \(\eta\) of 0.45 for silicon vacancy (SiV) centers in diamond. This advance in quantum emitter design holds promise for enhancing the efficiency and scalability of quantum information systems.

\section{Acknowledgements}
The authors thank Yuqin Duan and Matthew Trusheim for valuable discussions during the design process. This work was partially supported by the National Science Foundation (NSF) award number 1941583.


\bibliography{citations} 

\begin{thebibliography}{10}
\newcommand{\enquote}[1]{``#1''}

\bibitem{Ruf2021}
M.~Ruf, N.~H. Wan, H.~Choi, D.~Englund, and R.~Hanson, \enquote{{Quantum networks based on color centers in diamond},} {\protect\JournalTitle{Journal of Applied Physics}} \textbf{130}, 070901 (2021).

\bibitem{Pompili2021}
M.~Pompili, S.~L.~N. Hermans, S.~Baier, H.~K.~C. Beukers, P.~C. Humphreys, R.~N. Schouten, R.~F.~L. Vermeulen, M.~J. Tiggelman, L.~dos Santos~Martins, B.~Dirkse, S.~Wehner, and R.~Hanson, \enquote{Realization of a multinode quantum network of remote solid-state qubits,} {\protect\JournalTitle{Science}} \textbf{372}, 259--264 (2021).

\bibitem{Orphal2023}
L.~Orphal-Kobin, K.~Unterguggenberger, T.~Pregnolato, N.~Kemf, M.~Matalla, R.-S. Unger, I.~Ostermay, G.~Pieplow, and T.~Schr\"oder, \enquote{Optically coherent nitrogen-vacancy defect centers in diamond nanostructures,} {\protect\JournalTitle{Phys. Rev. X}} \textbf{13}, 011042 (2023).

\bibitem{duan21}
Y.~Duan, K.~C. Chen, D.~R. Englund, and M.~E. Trusheim, \enquote{A vertically-loaded diamond microdisk resonator spin-photon interface,} {\protect\JournalTitle{Opt. Express}} \textbf{29}, 43082--43090 (2021).

\bibitem{Humphreys2018-pj}
P.~C. Humphreys, N.~Kalb, J.~P.~J. Morits, R.~N. Schouten, R.~F.~L. Vermeulen, D.~J. Twitchen, M.~Markham, and R.~Hanson, \enquote{Deterministic delivery of remote entanglement on a quantum network,} {\protect\JournalTitle{Nature}} \textbf{558}, 268--273 (2018).

\bibitem{li15}
L.~Li, T.~Schr{\"o}der, E.~H. Chen, M.~Walsh, I.~Bayn, J.~Goldstein, O.~Gaathon, M.~E. Trusheim, M.~Lu, J.~Mower, M.~Cotlet, M.~L. Markham, D.~J. Twitchen, and D.~Englund, \enquote{Coherent spin control of a nanocavity-enhanced qubit in diamond,} {\protect\JournalTitle{Nature Communications}} \textbf{6}, 6173 (2015).

\bibitem{gould16}
M.~Gould, S.~Chakravarthi, I.~R. Christen, N.~Thomas, S.~Dadgostar, Y.~Song, M.~L. Lee, F.~Hatami, and K.-M.~C. Fu, \enquote{Large-scale gap-on-diamond integrated photonics platform for nv center-based quantum information,} {\protect\JournalTitle{J. Opt. Soc. Am. B}} \textbf{33}, B35--B42 (2016).

\bibitem{Bradac2019-hl}
C.~Bradac, W.~Gao, J.~Forneris, M.~E. Trusheim, and I.~Aharonovich, \enquote{Quantum nanophotonics with group {IV} defects in diamond,} {\protect\JournalTitle{Nature Communications}} \textbf{10}, 5625 (2019).

\bibitem{trusheim2020-mt}
M.~E. Trusheim, B.~Pingault, N.~H. Wan, M.~G\"undo\ifmmode~\breve{g}\else \u{g}\fi{}an, L.~De~Santis, R.~Debroux, D.~Gangloff, C.~Purser, K.~C. Chen, M.~Walsh, J.~J. Rose, J.~N. Becker, B.~Lienhard, E.~Bersin, I.~Paradeisanos, G.~Wang, D.~Lyzwa, A.~R.-P. Montblanch, G.~Malladi, H.~Bakhru, A.~C. Ferrari, I.~A. Walmsley, M.~Atat\"ure, and D.~Englund, \enquote{Transform-limited photons from a coherent tin-vacancy spin in diamond,} {\protect\JournalTitle{Phys. Rev. Lett.}} \textbf{124}, 023602 (2020).

\bibitem{Takou:21}
E.~Takou and S.~E. Economou, \enquote{Optical control protocols for high-fidelity spin rotations of single $\mathrm{Si}{\mathrm{v}}^{\ensuremath{-}}$ and $\mathrm{Sn}{\mathrm{v}}^{\ensuremath{-}}$ centers in diamond,} {\protect\JournalTitle{Phys. Rev. B}} \textbf{104}, 115302 (2021).

\bibitem{PhysRevLett.129.173603-Trusheim}
J.~Arjona~Mart\'{\i}nez, R.~A. Parker, K.~C. Chen, C.~M. Purser, L.~Li, C.~P. Michaels, A.~M. Stramma, R.~Debroux, I.~B. Harris, M.~Hayhurst~Appel, E.~C. Nichols, M.~E. Trusheim, D.~A. Gangloff, D.~Englund, and M.~Atat\"ure, \enquote{Photonic indistinguishability of the tin-vacancy center in nanostructured diamond,} {\protect\JournalTitle{Phys. Rev. Lett.}} \textbf{129}, 173603 (2022).

\bibitem{PhysRevX.13.031022-Rosenthal}
E.~I. Rosenthal, C.~P. Anderson, H.~C. Kleidermacher, A.~J. Stein, H.~Lee, J.~Grzesik, G.~Scuri, A.~E. Rugar, D.~Riedel, S.~Aghaeimeibodi, G.~H. Ahn, K.~Van~Gasse, and J.~Vu\ifmmode \check{c}\else \v{c}\fi{}kovi\ifmmode~\acute{c}\else \'{c}\fi{}, \enquote{Microwave spin control of a tin-vacancy qubit in diamond,} {\protect\JournalTitle{Phys. Rev. X}} \textbf{13}, 031022 (2023).

\bibitem{OSKOOI2010687}
A.~F. Oskooi, D.~Roundy, M.~Ibanescu, P.~Bermel, J.~Joannopoulos, and S.~G. Johnson, \enquote{Meep: A flexible free-software package for electromagnetic simulations by the fdtd method,} {\protect\JournalTitle{Computer Physics Communications}} \textbf{181}, 687--702 (2010).

\bibitem{Zhu:13}
J.~Zhu, X.~Cai, Y.~Chen, and S.~Yu, \enquote{Theoretical model for angular grating-based integrated optical vortex beam emitters,} {\protect\JournalTitle{Opt. Lett.}} \textbf{38}, 1343--1345 (2013).

\bibitem{mitchell19}
M.~Mitchell, D.~P. Lake, and P.~E. Barclay, \enquote{{Realizing $Q >$ 300 000 in diamond microdisks for optomechanics via etch optimization},} {\protect\JournalTitle{APL Photonics}} \textbf{4}, 016101 (2019).

\bibitem{xie2023}
J.~Xie, J.~Qian, T.~Wang, L.~Zhou, X.~Zang, L.~Chen, Y.~Zhu, and S.~Zhuang, \enquote{{Integrated terahertz vortex beam emitter for rotating target detection},} {\protect\JournalTitle{Advanced Photonics}} \textbf{5}, 066002 (2023).

\bibitem{john10}
J.~B. Schneider, \emph{Understanding the Finite-Difference Time-Domain Method} (www.eecs.wsu.edu/~schneidj/ufdtd, 2010).

\bibitem{Ulaby_Ravaioli_2023}
F.~T. Ulaby and U.~Ravaioli, \emph{Fundamentals of Applied Electromagnetics} (Pearson, 2023).

\bibitem{Lumerical}
{ANSYS}, \enquote{Lumerical,} {\protect\JournalTitle{https://www.lumerical.com/}} p. 2023 R2 (2023).

\end{thebibliography}

\end{document}